\documentclass[11pt]{article}
\usepackage{natbib,fullpage,amsmath,amssymb,graphicx,hyperref}
\newcommand{\m}[1]{#1}

\begin{document}
\title{Modeling homophily and stochastic equivalence in
       symmetric relational data}
\author{Peter D. Hoff \thanks{Departments of Statistics, Biostatistics and the
Center for Statistics and the Social Sciences,
University of Washington,
Seattle, Washington 98195-4322.
Web: \href{http://www.stat.washington.edu/hoff/}{\tt http://www.stat.washington.edu/hoff/}. This work was partially funded by NSF grant number 0631531.}
        }
\date{ \today }
\maketitle

\begin{abstract}
This article discusses a latent variable model for inference and prediction of 
symmetric relational data. 
  The model, 
based on the idea of the eigenvalue decomposition, represents the 
relationship between two nodes as the weighted inner-product of 
node-specific vectors of latent characteristics. This ``eigenmodel''
generalizes other popular latent variable models, such as 
latent class and distance models: It is shown  
 mathematically that any 
latent class or distance  model  has a representation as an 
 eigenmodel, but not vice-versa. 
The practical implications of this are examined in the context of 
three real datasets, for which 
the eigenmodel has as good or better out-of-sample predictive 
performance than the other two models.

\vspace{.2in}
\noindent {\it Some key words}:
Factor analysis, latent class,  
Markov chain Monte Carlo,
social network. 
\end{abstract}


\section{Introduction}
Let $\{y_{i,j} : 1\leq i< j \leq n \}$ denote  
data  measured
on pairs of a set of $n$ objects or nodes. 
The examples considered in this article include 
friendships  among people, 
associations among words and
interactions among proteins.
Such measurements are often
represented by a sociomatrix $\m Y$, which is a symmetric $n\times n$  matrix with an
undefined diagonal. 
One of the goals of relational data analysis is to describe the variation 
among the entries of $\m Y$, 
as well as any potential covariation of $\m Y$ with 
observed explanatory variables $X=\{ \m x_{i,j}, 1\leq i< j\leq n\}$. 

To this end, a variety of statistical  models have been developed that 
describe $y_{i,j}$ 
as some function of 
node-specific latent variables $u_i$ and $u_j$ and 
a linear predictor $\beta^T\m x_{i,j}$. 
In such formulations, 
$\{u_1,\ldots, u_n\}$ represent across-node variation in the $y_{i,j}$'s
and 
$\beta$ represents covariation of the $y_{i,j}$'s  with the $\m x_{i,j}$'s. 
For example,  \citet{nowicki_snijders_2001} present a model in 
which each node $i$ is assumed to belong to an unobserved latent 
class $u_i$, and a probability distribution describes the  
relationships between each pair of classes (see 
 \citet{kemp_griffiths_tenenbaum_2004} and
\citet{airoldi_2005} for recent extensions of this approach). 
 Such a model captures 
{\em stochastic equivalence}, a type of pattern often seen in 
network data in which the nodes can be divided into groups 
such that members of the same group have similar patterns of 
relationships. 

An alternative approach  to 
representing across-node variation is based on the idea of  {\em homophily},
in which  the relationships between nodes with similar characteristics 
are stronger than the 
relationships between nodes having different characteristics. 
Homophily  provides an explanation to  data patterns often seen in social networks,  such as
transitivity (``a friend of a friend is a friend''),
balance (``the enemy of my friend is an enemy'') and the existence
of cohesive subgroups of nodes.
In order to represent such patterns, 
\citet{hoff_raftery_handcock_2002} present a
model in which the conditional mean of $y_{i,j}$ 
is a function of $\beta'\m x_{i,j} - |\m u_i - \m u_j|$, 
where $\{\m u_1,\ldots, \m u_n\}$ are vectors of unobserved, latent characteristics
in a Euclidean space. 
In the context of binary relational data, such a model predicts the
existence of
more transitive triples, or ``triangles,''  than would be seen under 
a random allocation of edges among pairs of nodes.
An important assumption of this model is that 
two nodes with a strong relationship between them 
are  also similar to each other in terms of 
how they relate to other nodes:
A strong relationship between $i$ and $j$ suggests
$|u_i- u_j|$  is small, but this further implies that
 $|u_i -u_k| \approx | u_j -u_k|$,  and so nodes $i$ and $j$ are assumed
to have similar relationships to other  nodes.

\begin{figure}
\centerline{\includegraphics[height=2.25in]{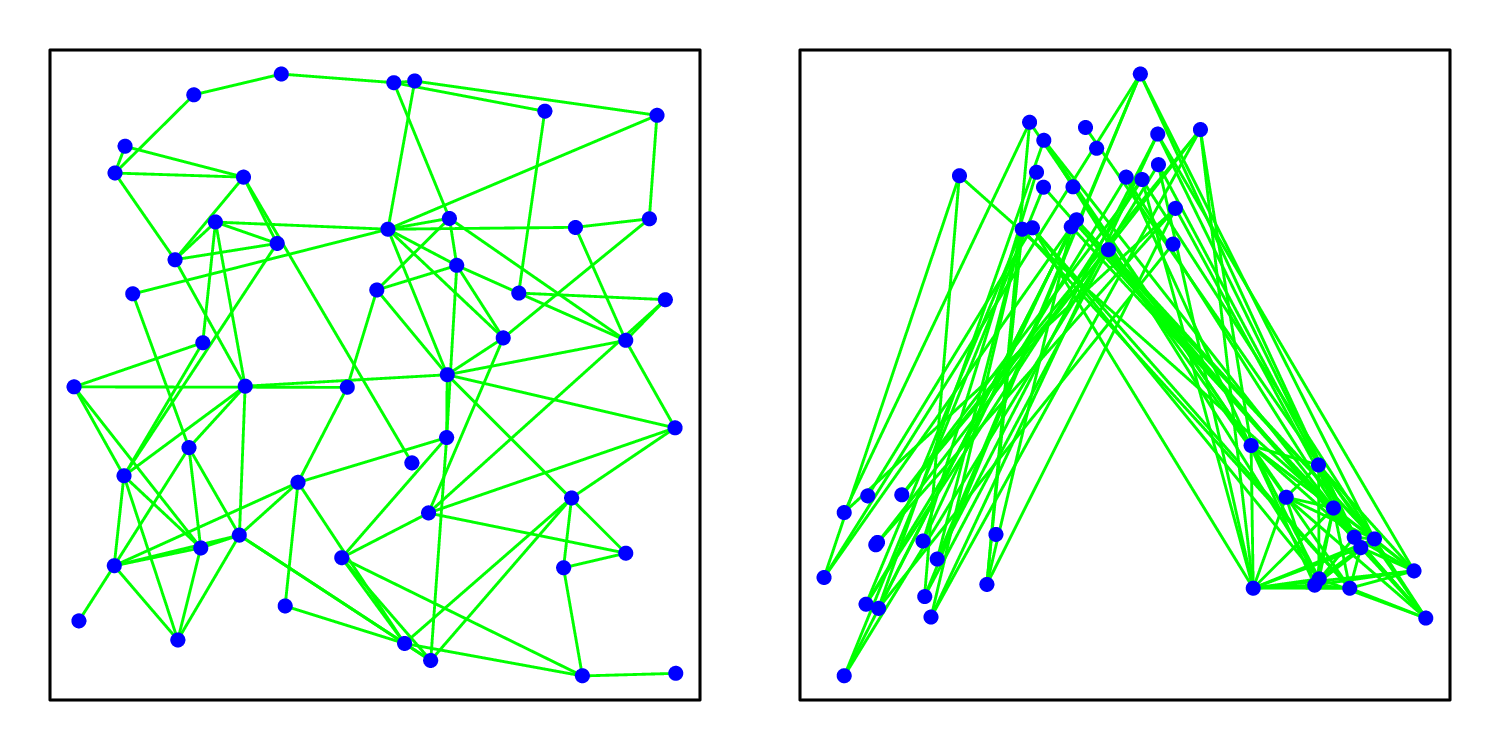}}
\caption{Networks exhibiting  homophily (left panel) and stochastic
 equivalence (right panel). }
\label{fig:exdata_simu}
\end{figure}

The latent class model of \citet{nowicki_snijders_2001} and the 
latent distance model of \citet{hoff_raftery_handcock_2002} 
are able to identify,  respectively, 
classes of nodes with similar roles, and the locational properties 
of the nodes. These two items are 
perhaps the two primary features of interest in social network and 
relational data analysis. For example, discussion of these concepts  makes up more
than half of the 734 pages of main text in 
\citet{wasserman_faust_1994}.
However, a model that can represent one feature may not be 
able to represent the other: Consider the two 
graphs in Figure \ref{fig:exdata_simu}.  The graph on the
left  displays a large degree of transitivity, and can  
be well-represented by the latent distance model with a set of 
vectors $\{ u_1,\ldots, u_n\}$ in two-dimensional space, in which 
the probability of an edge  between $i$ and $j$ is decreasing in  $|u_i-u_j|$. 
In contrast, representation of the graph by a latent class model 
would require a large number of classes,  none of which would be 
particularly cohesive or distinguishable from the others. 
The second panel of Figure \ref{fig:exdata_simu}  displays
a network involving three classes
of stochastically equivalent nodes, two of which (say $A$ and $B$) have only across-class ties, and one ($C$) that has 
both within- and  across-class ties. This graph is well-represented
by a  latent class model in which edges occur with high probability 
between pairs having one member in each of $A$ and $B$ or in $B$ and $C$, 
and among pairs having both members in $C$
(in models of stochastic equivalence,  nodes within each class are not 
differentiated).  
In contrast, representation of this type of graph with a latent 
distance model would require the dimension of the latent characteristics
to be on the order of the class membership sizes. 

Many real networks exhibit combinations of structural equivalence and
homophily 
in varying degrees.
In these situations, use of either the latent class or  distance model
would only be representing part of the network structure.
The goal of this paper is to show that a simple statistical model based on 
the eigenvalue decomposition can generalize the latent class and distance 
models: 
 Just as any symmetric matrix can be approximated 
with a subset of its largest eigenvalues and corresponding eigenvectors, 
the variation in a sociomatrix can be represented by  modeling 
 $y_{i,j}$ as a function of  $\beta'\m x_{i,j} + \m u_i ^T\Lambda \m u_j$, 
where $\{u_1,\ldots, u_n\}$ are node-specific factors and $\Lambda$ 
is a diagonal matrix.
In this article, we show mathematically and by example 
how this eigenmodel  can represent both  stochastic equivalence and homophily in symmetric relational data, and thus is more general than the 
 other two latent variable models. 

The next section motivates the use of latent variables models for relational 
data, and shows 
mathematically  that the eigenmodel
generalizes the latent class and distance models in the
sense that it
can compactly represent  the same network features as these
other models but not vice-versa.
Section 3 compares the out-of-sample predictive performance of these
 three models on three different datasets: 
a social network of 12th graders; 
a relational dataset on word association counts from the 
first chapter of Genesis; 
and a dataset 
on protein-protein interactions. 
The first two networks exhibit latent homophily   and stochastic equivalence
respectively, whereas the third shows both to some degree. 
In support of the theoretical 
results of Section 2, the latent distance and class models 
perform well for the first and second datasets respectively, whereas the 
 eigenmodel performs well for all three. 
Section 4 summarizes the results and discusses some extensions. 


\section{Latent variable modeling of relational  data}

\subsection{Justification of latent variable modeling}
The use of probabilistic latent variable models for the representation of relational data 
can be motivated 
in a natural way: 
For  undirected data without covariate information, 
symmetry suggests that any probability model we consider
should treat 
the nodes  as being exchangeable,
so that
\[ \Pr (\{ y_{i,j} : 1\leq i < j\leq n \}   \in A ) = 
    \Pr (\{ y_{\pi i,\pi j} : 1\leq i < j\leq n \}   \in A ) \]
for any permutation $\pi$ of the integers $\{1,\ldots, n\}$ and
any set of sociomatrices $A$.
Results of \citet{hoover_1982} and \citet[chap. 14]{aldous_1985}
show that  if a model 
satisfies the above exchangeability condition for each 
integer $n$, then it can be written as a latent variable
model of the form
\begin{equation}
 y_{i,j} = h(\mu,u_i, u_j, \epsilon_{i,j} ) 
\label{eq:decomp}
\end{equation}
for i.i.d.\ latent variables $\{u_1,\ldots, u_n\}$,
i.i.d.\ pair-specific effects $\{\epsilon_{i,j}: 1\leq i<j\leq n\}$
and  some function $h$ that is symmetric in its second and third
arguments. This result is very general - it says that
any statistical model for a sociomatrix in which the nodes
are exchangeable can be written as a latent variable model.

Difference choices of $h$ lead to different models for $y$. 
A general probit model for binary network data can be put in the 
form of (\ref{eq:decomp}) as follows:
\begin{eqnarray}
\{\epsilon_{i,j}: 1\leq i<j\leq n\}& \sim  & \mbox{i.i.d.\ normal}(0,1)
   \nonumber \\
\{u_1,\ldots, u_n\} &\sim & \mbox{i.i.d.} \  f(u|\psi)  \nonumber \\
y_{i,j} =h(\mu,u_i,u_j,\epsilon_{i,j})&=& \delta_{ (0,\infty) } (\mu+ \alpha(u_i,u_j) + \epsilon_{i,j} ),  \nonumber
\end{eqnarray}
where $\mu$ and $\psi$ are parameters to be estimated, and $\alpha$ is 
a symmetric function, also potentially involving parameters
to be estimated. Covariation between $Y$ 
and an array of predictor variables $X$ can be represented by adding  
a linear predictor $\beta^Tx_{i,j}$ to $\mu$.  Finally, 
integrating over $\epsilon_{i,j}$ we obtain 
$\Pr(y_{i,j}=1|x_{i,j},u_i,u_j) =  \Phi[\mu+ \beta^Tx_{i,j}  + \alpha(u_i,u_j)]$. 
Since the $\epsilon_{i,j}$'s can be assumed to be independent, the 
conditional probability of $Y$ given  $X$ and $\{u_1,\ldots, u_n\}$
can be expressed as 
\begin{eqnarray}
\Pr(y_{i,j}=1 |x_{i,j} , u_i,u_j ) 
 \equiv  \theta_{i,j} 
 &=&    \Phi[ \mu +\beta^T x_{i,j}  + \alpha(u_i,u_j)]\label{eq:probit}\\
 \Pr(Y|X,u_1,\ldots, u_n) &=& \prod_{i<j}  \theta^{y_{i,j}}_{i,j}  
 (1-\theta_{i,j})^{y_{i,j}}  \nonumber
\end{eqnarray}
Many relational datasets have ordinal, non-binary 
measurements (for example, the word association data in Section 3.2). 
Rather than ``thresholding'' the data to force it to be binary, 
we can make use of the full information in the data with an
ordered probit version of (\ref{eq:probit}):
\begin{eqnarray*}
\Pr(y_{i,j} = y |x_{i,j} , u_i,u_j )\equiv  \theta_{i,j}^{(y)} &=& \Phi[ \mu_{y} +\beta^T x_{i,j}  + \alpha(u_i,u_j)]- \Phi[ \mu_{y+1} +\beta^T x_{i,j}  + \alpha(u_i,u_j)] \\
 \Pr(Y|X,u_1,\ldots, u_n) &=& \prod_{i<j}  \theta^{(y_{i,j})}_{i,j}  , 
\end{eqnarray*}
where $\{ \mu_y \}$ are parameters to be estimated for all 
but the lowest value $y$ in the sample space.

\subsection{Effects of nodal variation}
The latent variable models described in the Introduction correspond to 
different choices for the symmetric function $\alpha$:
\begin{description}
\item{Latent class model:}   \
\begin{itemize}
\item[] $\alpha(u_i,u_j) = m_{u_i,u_j}$
\item[] $u_i \in \{1,\ldots, K\}, \ i\in \{1,\ldots, n\}$
\item[] $M$ a $K \times K $ symmetric matrix
\end{itemize}
\item{Latent distance model:} \ 
\begin{itemize}
\item[] $\alpha (u_i,u_j) =  - |u_i - u_j|$  
\item[] $u_i \in \mathbb R^K, \ i\in \{1,\ldots, n\}$
\end{itemize}
\item{Latent eigenmodel:} \ 
\begin{itemize}
\item[]  $\alpha(u_i,u_j) = u_i^T\Lambda u_j$
\item[] $u_i \in \mathbb R^K, \ i\in \{1,\ldots, n\}$
\item[]  $\Lambda$ a
$K\times K$ diagonal matrix.
\end{itemize}
\end{description}
Interpretations of the latent class and distance models were 
given in the Introduction. An interpretation of the latent  
eigenmodel is that each node $i$ has a vector of unobserved characteristics 
$u_i=\{ u_{i,1},\ldots, u_{i,K}\}$, and that similar 
values of $u_{i,k}$ and $u_{j,k}$ will contribute positively or 
negatively to the relationship between $i$ and $j$, depending on 
whether $\lambda_{k}>0$ or $\lambda_{k}<0$. 
In this way, the model can represent 
both positive or negative homophily in varying degrees, and 
stochastically 
 equivalent nodes (nodes with the same or similar latent vectors) 
may or may not have strong relationships with one another. 

We now show that the eigenmodel generalizes the latent class and 
distance models:
Let $\mathcal S_n$ be the set of $n\times n$ sociomatrices, 
and let 
\begin{eqnarray*}
\mathcal C_K &=& \{ C\in \mathcal S_n: c_{i,j} = m_{u_i,u_j},  \
    u_i \in \{1,\ldots, K\}, 
        \mbox{ $M$ a $K \times K $ symmetric matrix}\}; \\
\mathcal D_K &=& \{ D\in \mathcal S_n: d_{i,j} = -|u_i-u_j|,  \ 
       u_i\in \mathbb R^K 
  \};  \\
\mathcal E_K &=& \{ E\in \mathcal S_n: e_{i,j} = u_i^T\Lambda u_j, \  
     u_i \in \mathbb R^K, \ 
     \mbox{ $\Lambda$ a $K \times K $ diagonal matrix}\} .
\end{eqnarray*}
In other words, $\mathcal C_K$ is the set  of possible values 
of $\{ \alpha(u_i,u_j), 1\leq i<j\leq n\}$ under a $K$-dimensional latent class model, 
and similarly for $\mathcal D_K$ and $\mathcal E_K$. 

\paragraph{$\mathcal E_K$ generalizes $\mathcal C_K$:} 
Let $C\in \mathcal C_K$ and let $\tilde C$ be a completion of 
$C$ obtained by setting $c_{i,i} = m_{u_i,u_i}$. There are 
at most $K$ unique rows of $\tilde C$ and so $\tilde C$ is 
of rank $K$ at most. 
Since the set $\mathcal E_K$ contains all sociomatrices that can be completed 
 as a rank-$K$ matrix, we have $\mathcal C_K \subseteq \mathcal E_K$. 
Since  $\mathcal E_K$ includes matrices with $n$ unique rows, 
 $\mathcal C_K \subset \mathcal E_K$ unless $K\geq n$ in which 
case the two sets are equal. 

\paragraph{$\mathcal E_{K+1}$ weakly generalizes $\mathcal D_K$:}  
Let $D \in \mathcal D_K$. 
Such a (negative) distance matrix will generally be of full rank, 
in which case it
 cannot be represented exactly  by an $E\in \mathcal E_K$ for $K<n$. 
However, what is critical from a modeling perspective is whether or not 
the \emph{order} of the entries of each $D$ can be matched by the order of the entries 
of an $E$. 
This is because the probit and ordered probit model we are considering 
include threshold variables $\{ \mu_y: y\in \mathcal Y\}$ which 
can be adjusted to accommodate monotone transformations of 
$\alpha(u_i,u_j)$. 
With this in mind, note that the
 matrix of \emph{squared} distances among a set of $K$-dimensional vectors 
$\{z_1,\ldots, z_n\}$  is a monotonic transformation 
of the distances, is of rank $K+2$ or less 
(as $D^2 =  [ z_1'z_1,\ldots, z_n'z_n]^T {\boldmath 1}^T +  
    {\boldmath 1} [ z_1'z_1,\ldots, z_n'z_n]   -2 Z Z^T $)
and so is in $\mathcal E_{K+2}$. 
Furthermore, 
letting
$u_{i}  = ( z_i , \sqrt{r^2 -z_i^Tz_i}) \in \mathbb R^{K+1} $ for each 
$i\in\{1,\ldots, n\}$, we have $u_i'u_j = 
  z_i'z_j + \sqrt{ (r^2-|u_i|^2)(r^2-|u_j|^2) }$. For large $r$ this 
is approximately $r^2 - |z_i-z_j|^2/2$, which is an increasing 
function of the negative distance $d_{i,j}$. For large enough $r$
the numerical order of the entries of this  $E\in \mathcal E_{K+1}$ 
is the same as that of $D\in \mathcal D_K$.

\paragraph{$\mathcal D_K$ does not weakly generalize $\mathcal E_1$:}
Consider $E\in \mathcal E_1$ generated by $\Lambda=1$,  $u_1 =1$  and 
$u_i =r <1 $ for $i>1$. Then $r=e_{1,i_1} =e_{1,i_2}  > e_{i_1,i_2} =r^2$ 
for all $i_1,i_2\neq 1$. For which $K$ is such an ordering of the elements
of 
$D \in \mathcal D_K$ possible? If $K=1$ then such an ordering is possible 
only if $n=3$. For $K=2$ such an ordering is possible for $n\leq 6$. This is 
because the \emph{kissing number} in $\mathbb R^2$, 
or the number of non-overlapping spheres of unit radius that 
can simultaneously touch a central sphere of unit radius, 
is 6. If we put node 1 at the center of the central 
sphere, and 6 nodes at the centers of the 6 kissing spheres, then 
we have $d_{1,i_1} =d_{1,i_2} = d_{i_1,i_2}$    
for all $i_1,i_2\neq 1$. We can only have 
$d_{1,i_1} =d_{1,i_2} > d_{i_1,i_2}$  if we remove one of the non-central
 spheres
to allow for more room between those remaining, leaving 
one central 
sphere plus five kissing spheres
for a total of $n=6$. 
Increasing $n$ increases the 
necessary 
dimension of the Euclidean space, and so for any $K$ there are $n$ and $E\in \mathcal E_1$ 
that have entry orderings that cannot be matched by those of  any $D\in \mathcal D_K$.

A 
less general positive semi-definite version of the eigenmodel 
has been studied by
\citet{hoff_2005}, in which $\Lambda$ was taken to be the identity matrix. 
Such a model can weakly generalize a distance model, but cannot generalize 
a latent class model, 
 as the eigenvalues of a latent class model could 
be negative.




\section{Model comparison on three different datasets}

\subsection{Parameter estimation}
Bayesian parameter estimation for the three models under consideration 
can be achieved via Markov chain Monte Carlo (MCMC) algorithms, 
in which posterior distributions for the unknown quantities are 
approximated with empirical distributions of samples 
from a Markov chain. 
For these algorithms, it is useful to 
formulate 
the probit 
models described in Section 2.1 in terms of an additional 
latent variable 
 $z_{i,j} \sim $ 
normal$[\beta'x_{i,j}+ \alpha(u_i,u_j)]$,  for which 
$y_{i,j} = y $ if $\mu_y < z_{i,j} < \mu_{y+1}$.  
Using 
conjugate prior distributions where possible, the MCMC algorithms proceed
by generating a new state  $\phi^{(s+1)} = 
 \{ Z^{(s+1)}, \mu^{(s+1)} ,  \beta^{(s+1)}, u_1^{(s+1)},\ldots, u_n^{(s+1)}\}$
from a current state $\phi^{(s)}$ as follows:
\begin{enumerate}
\item 
For each $\{i,j\}$, sample $z_{i,j}$ from 
     its (constrained normal) full 
conditional distribution. 
\item  For each $y\in \mathcal Y$, 
      sample $\mu_y$ from its (normal) full 
      conditional distribution. 
\item  
Sample $\beta$ from its (multivariate normal) full conditional distribution. 
\item Sample $u_1,\ldots, u_n$ and their associated parameters: 
 \begin{itemize}
\item For the latent  distance model, propose and accept or reject new values of the $u_i$'s with the  Metropolis algorithm,  and then sample 
  the population variances of the $u_i$'s from their (inverse-gamma) full conditional 
      distributions. 
\item For the latent class model, update each class variable $u_i$ 
      from its (multinomial) conditional distribution given current values of $\m Z, \{u_j:j\neq i\}$ and 
      the variance of the elements of $M$ (but marginally over $M$ to improve mixing). Then sample the elements of
    $M$ from their
      (normal) full conditional distributions  and the variance 
     of the entries of $M$ from its (inverse-gamma)  full conditional distribution. 
\item For the latent vector model, sample each $u_i$ from its (multivariate normal) full 
conditional distribution, 
 sample the mean of the $u_i$'s from their  (normal)
  full conditional distributions, 
and then sample $\Lambda$ from 
its (multivariate normal) full conditional distribution. 
\end{itemize}
\end{enumerate}
To facilitate comparison across models, we used
prior distributions in which the level of prior variability
in $\alpha(u_i,u_j)$ was similar across the three different models. 
An {\sf R} package that implements the MCMC  is available at 
\href{http://cran.r-project.org/src/contrib/Descriptions/eigenmodel.html}{\small \tt cran.r-project.org/src/contrib/Descriptions/eigenmodel.html}.


\subsection{Cross validation}
To compare the performance of these three different models
we evaluated their out-of-sample predictive performance 
under a range of dimensions ($K\in \{3,5,10\}$) and 
on three different datasets exhibiting varying combinations 
of homophily and stochastic equivalence. For each 
combination of dataset, dimension and model 
we performed a five-fold cross validation 
experiment as follows: 
\begin{enumerate}
\item Randomly divide the $n \choose 2$ data values into 5 sets
     of roughly equal size, letting $s_{i,j}$ be the set to which
      pair $\{i,j\}$ is assigned. 
\item For each  $s\in \{1,\ldots, 5\}$:
\begin{enumerate}
\item Obtain posterior distributions of the model parameter
conditional on  $\{ y_{i,j}: s_{i,j}\neq s\}$, 
 the data on pairs not in set $s$. 
\item  For pairs $\{ k,l\} $ in set  $s$, let $\hat y_{k,l} = E[y_{k,l} |  
     \{y_{i,j} : s_{i,j}\neq s\} ]$, the posterior predictive mean 
    of $y_{k,l}$ obtained using data not in set $s$. 
\end{enumerate}
\end{enumerate}
This procedure generates a  sociomatrix $\hat Y$, in which 
each entry $\hat y_{i,j}$ represents a predicted value obtained 
from using  a subset of the data that does not 
include $y_{i,j}$. Thus $\hat Y$ is a sociomatrix of out-of-sample predictions 
of 
the observed data $Y$. 

\begin{table}[t]
\caption{Cross validation results and area under the ROC curves.}
\label{roc}
\begin{center}
\begin{tabular}{r|ccc|ccc|ccc}
$K$ & \multicolumn{3}{c}{\bf Add health}  &\multicolumn{3}{c}{\bf Genesis}
 &\multicolumn{3}{c}{\bf Protein interaction}   \\
 & dist & class & eigen&  dist & class & eigen  & dist & class & eigen  \\ \hline
3 & 0.82 & 0.64 & 0.75 & 0.62 & 0.82 & 0.82 &  0.83  & 0.79 & 0.88  \\
5 & 0.81  & 0.70 & 0.78 &  0.66 & 0.82 & 0.82 & 0.84 & 0.84 & 0.90   \\
10 & 0.76 & 0.69 & 0.80 & 0.74 & 0.82 & 0.82 & 0.85 & 0.86 & 0.90
\end{tabular}
\end{center}
\end{table}

\subsection{Adolescent Health  social network}
\begin{figure}[t]
\centerline{\includegraphics[height=2.75in]{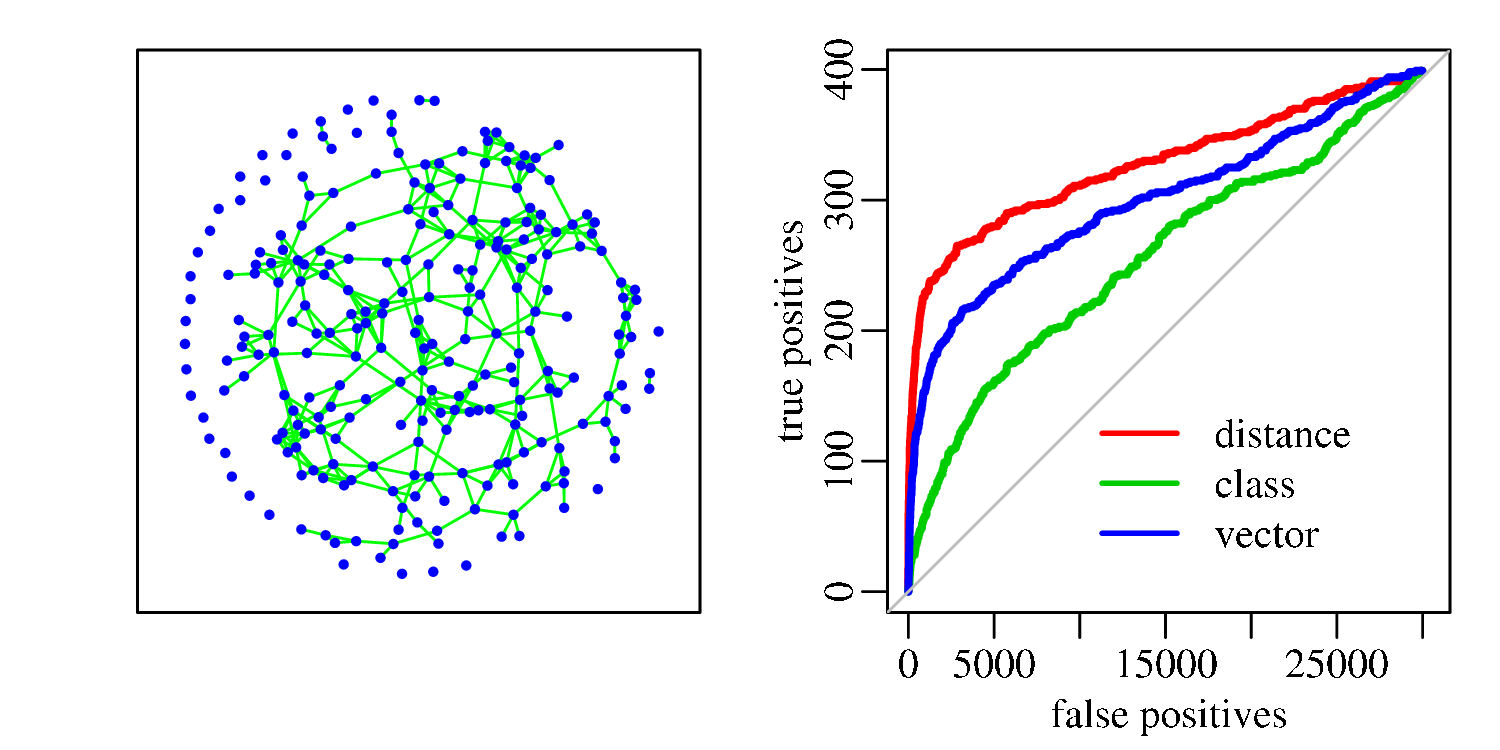}}
\caption{Social network data and unscaled ROC curves for the $K=3$ models.}
\label{fig:adhealth}  
\end{figure}

The first dataset records  friendship ties among 
247 12th-graders, obtained from the
National Longitudinal
Study of Adolescent Health 
(\href{http://www.cpc.unc.edu/projects/addhealth}{\small \tt www.cpc.unc.edu/projects/addhealth}). 
For these data, $y_{i,j} = 1 $ or $0$ depending on whether or not 
there is a close friendship tie between student $i$ and $j$ (as reported by 
either $i$ or $j$). 
These data are 
 represented as an undirected graph in the first panel of 
Figure \ref{fig:adhealth}.  
Like many social networks, these data exhibit a good deal of transitivity. 
It is therefore not surprising that 
the best performing models considered (in terms of area under the 
ROC curve, given in Table 1)
 are the distance models, with the eigenmodels close 
behind. In contrast, the latent class models perform poorly, 
and the results suggest that increasing $K$ for this model 
would not improve its performance. 


\begin{figure}
\centerline{\includegraphics[height=2.75in]{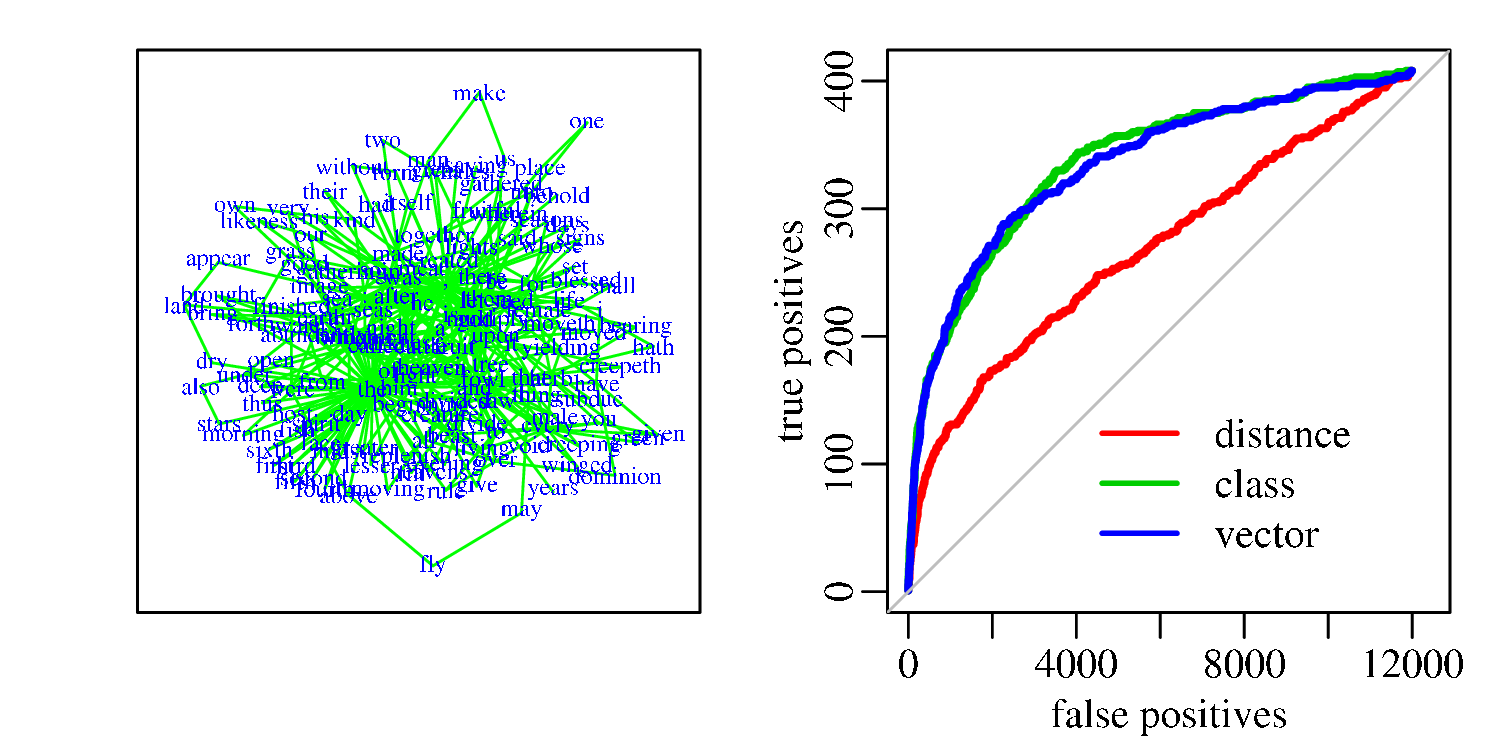}}
\caption{Relational text data from Genesis and unscaled ROC curves for the
   $K=3$ models.}
\label{fig:gen}
\end{figure}

\subsection{Word neighbors in Genesis}
The second dataset we consider is derived from 
 word  and punctuation counts
in the first chapter of the King James version of
Genesis (\href{http://www.gutenberg.org/dirs/etext05/bib0110.txt}{\small \tt www.gutenberg.org/dirs/etext05/bib0110.txt}). There are 158 unique words and 
punctuation marks in this 
chapter, and for our example we take
$y_{i,j}$ to be the number of times that word $i$ and word $j$ appear 
next to each other (a model extension, appropriate for an asymmetric 
version of this dataset, is discussed in the next section). 
These data can be viewed as a graph with weighted edges, the unweighted version of which is shown 
in the first panel of Figure \ref{fig:gen}. 
The lack of a clear spatial representation of these data is 
 not unexpected, as text  data such as these 
do not have groups of words with strong within-group connections, 
nor do they display much homophily: a given noun may appear quite frequently 
next to two different verbs, but these verbs will not appear next to 
each other. 
A better description of these data might be that there are classes of 
words,
and  connections occur between words of different classes. 
The cross validation results support this claim, 
in that the latent class model performs much better than the distance model 
on these data, as seen in the second panel of Figure \ref{fig:gen}
 and in 
Table 1. 
As discussed in the previous section, the 
eigenmodel generalizes the latent class model and performs equally well. 
We note that parameter estimates for these data were obtained using the 
ordered probit versions of the models (as the data are not binary), but 
the out-of-sample predictive performance was evaluated based on 
each model's ability to predict a non-zero relationship.

\subsection{Protein-protein interaction data}
\begin{figure}
\centerline{\includegraphics[height=2.75in]{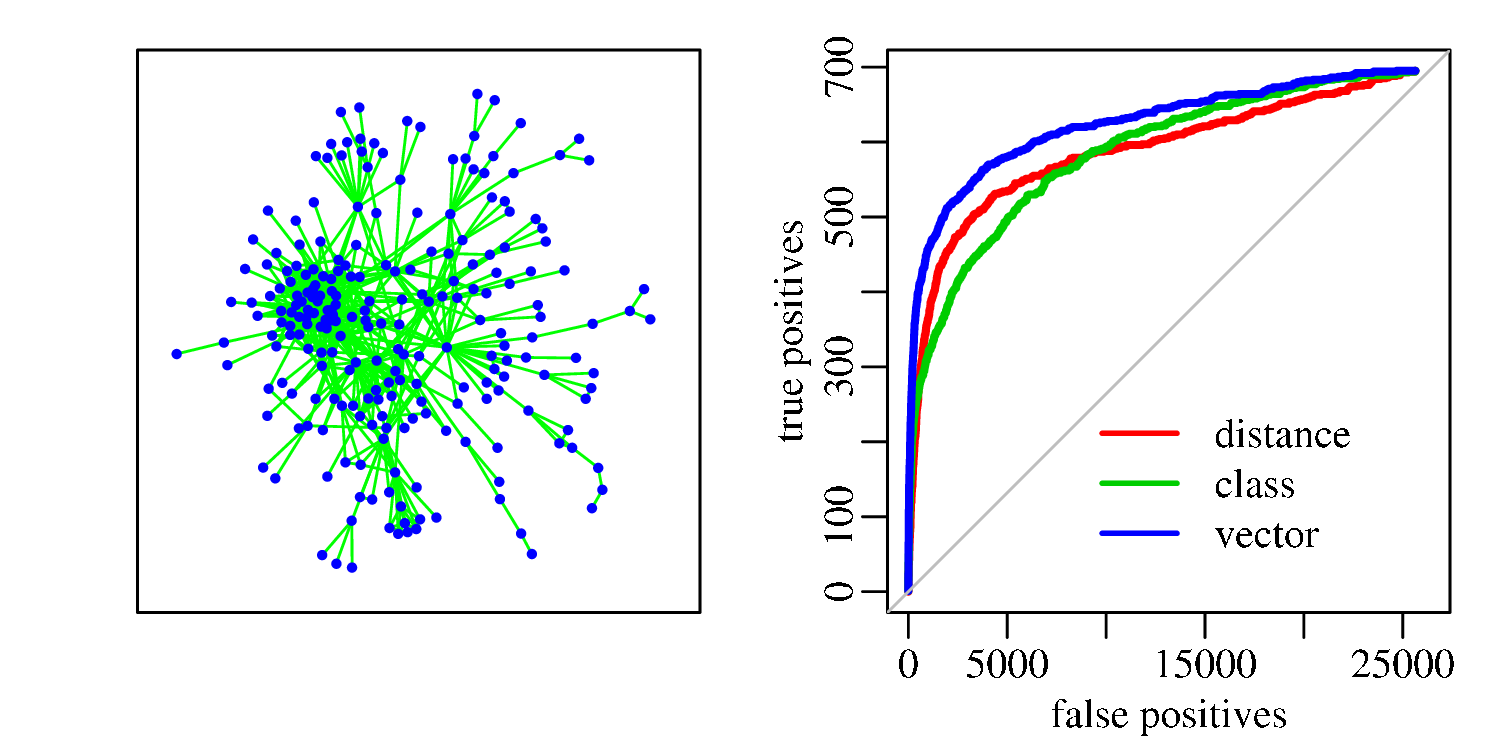}}
\caption{ Protein-protein interaction data and unscaled ROC curves
     for the $K=3$ models.}
\label{fig:pro}
\end{figure}

Our last example is the protein-protein interaction data of \citet{butland_2005}, 
 in which $y_{i,j}= 1 $ if proteins $i$ and $j$ bind and $y_{i,j}=0$ otherwise.
We analyze the large connected component of this graph, which includes 230
proteins and is displayed in the first panel of \ref{fig:pro}. 
This graph indicates patterns of both stochastic equivalence and 
homophily: Some nodes could be described as ``hubs'', connecting to 
many  other nodes which in turn do not connect to each other. Such structure
is better represented by a latent class model than a distance model. 
However, most nodes connecting to hubs generally connect to only one 
hub, which is a feature that is hard to represent with a small 
number of latent classes. To represent this structure well, we would 
need two latent classes per hub, one for the hub itself and one for the 
nodes connecting to the hub. 
Furthermore, the core of the network (the nodes with more than two connections)
displays a good degree of homophily in the form of transitive triads, 
a feature which is easiest to represent with a distance model. 
The eigenmodel is able to capture both of these data features
and performs better than the other two models in 
terms of out-of-sample predictive performance. In fact, the $K=3$ eigenmodel 
performs better than the other two models for any value of $K$ considered. 

\section{Discussion}
Latent distance and latent class models provide concise, easily interpreted
 descriptions of social 
networks and relational data. However, neither of these models will provide
a complete picture of relational data that exhibit degrees of both 
homophily and stochastic equivalence. In contrast, we have shown that 
a latent eigenmodel is able to represent datasets with either or both 
of these data patterns.  This is due to the fact that the eigenmodel
provides an unrestricted  low-rank approximation to the sociomatrix, and is 
therefore 
 able to represent a wide array of patterns in the data. 

The concept behind the eigenmodel is the familiar eigenvalue decomposition
of a symmetric matrix. The analogue for directed networks or rectangular
matrix data would be a model based on the singular value decomposition, 
in which data $y_{i,j}$ could be modeled  as depending on $u_i^T D v_j$, 
where $u_i$ and $v_j$ represent vectors of
 latent row and column effects respectively. 
Statistical inference using the singular value decomposition 
for Gaussian 
data is  straightforward. 
A model-based version of the approach for binary and other 
non-Gaussian relational datasets 
could be implemented 
using the ordered probit model discussed in this paper.

\subsubsection*{Acknowledgment}
This work was partially funded by NSF grant number 0631531.

\bibliographystyle{plainnat}
\bibliography{hoff_hose}

\begin{thebibliography}{9}
\providecommand{\natexlab}[1]{#1}
\providecommand{\url}[1]{\texttt{#1}}
\expandafter\ifx\csname urlstyle\endcsname\relax
  \providecommand{\doi}[1]{doi: #1}\else
  \providecommand{\doi}{doi: \begingroup \urlstyle{rm}\Url}\fi

\bibitem[Airoldi et~al.(2005)Airoldi, Blei, Xing, and Fienberg]{airoldi_2005}
Edoardo Airoldi, David Blei, Eric Xing, and Stephen Fienberg.
\newblock A latent mixed membership model for relational data.
\newblock In \emph{LinkKDD '05: Proceedings of the 3rd international workshop
  on Link discovery}, pages 82--89, New York, NY, USA, 2005. ACM Press.
\newblock ISBN 1-59593-215-1.
\newblock \doi{http://doi.acm.org/10.1145/1134271.1134283}.

\bibitem[Aldous(1985)]{aldous_1985}
David~J. Aldous.
\newblock Exchangeability and related topics.
\newblock In \emph{\'Ecole d'\'et\'e de probabilit\'es de Saint-Flour,
  XIII---1983}, volume 1117 of \emph{Lecture Notes in Math.}, pages 1--198.
  Springer, Berlin, 1985.

\bibitem[Butland et~al.(2005)Butland, Peregrin-Alvarez, Li, Yang, Yang,
  Canadien, Starostine, Richards, Beattie, Krogan, Davey, Parkinson,
  Greenblatt, and Emili]{butland_2005}
G.~Butland, J.~M. Peregrin-Alvarez, J.~Li, W.~Yang, X.~Yang, V.~Canadien,
  A.~Starostine, D.~Richards, B.~Beattie, N.~Krogan, M.~Davey, J.~Parkinson,
  J.~Greenblatt, and A.~Emili.
\newblock Interaction network containing conserved and essential protein
  complexes in escherichia coli.
\newblock \emph{Nature}, 433:\penalty0 531--537, 2005.

\bibitem[Hoff(2005)]{hoff_2005}
Peter~D. Hoff.
\newblock Bilinear mixed-effects models for dyadic data.
\newblock \emph{J. Amer. Statist. Assoc.}, 100\penalty0 (469):\penalty0
  286--295, 2005.
\newblock ISSN 0162-1459.

\bibitem[Hoff et~al.(2002)Hoff, Raftery, and
  Handcock]{hoff_raftery_handcock_2002}
Peter~D. Hoff, Adrian~E. Raftery, and Mark~S. Handcock.
\newblock Latent space approaches to social network analysis.
\newblock \emph{J. Amer. Statist. Assoc.}, 97\penalty0 (460):\penalty0
  1090--1098, 2002.
\newblock ISSN 0162-1459.

\bibitem[Hoover(1982)]{hoover_1982}
D.~N. Hoover.
\newblock Row-column exchangeability and a generalized model for probability.
\newblock In \emph{Exchangeability in probability and statistics (Rome, 1981)},
  pages 281--291. North-Holland, Amsterdam, 1982.

\bibitem[Kemp et~al.(2004)Kemp, Griffiths, and
  Tenenbaum]{kemp_griffiths_tenenbaum_2004}
Charles Kemp, Thomas~L. Griffiths, and Joshua~B. Tenenbaum.
\newblock Discovering latent classes in relational data.
\newblock AI Memo 2004-019, Massachusetts Institute of Technology, 2004.

\bibitem[Nowicki and Snijders(2001)]{nowicki_snijders_2001}
Krzysztof Nowicki and Tom A.~B. Snijders.
\newblock Estimation and prediction for stochastic blockstructures.
\newblock \emph{J. Amer. Statist. Assoc.}, 96\penalty0 (455):\penalty0
  1077--1087, 2001.
\newblock ISSN 0162-1459.

\bibitem[Wasserman and Faust(1994)]{wasserman_faust_1994}
Stanley Wasserman and Katherine Faust.
\newblock \emph{Social Network Analysis: {M}ethods and Applications}.
\newblock Cambridge University Press, Cambridge, 1994.

\end{thebibliography}

\end{document}